\pdfoutput=1
\documentclass[aps,pre,reprint,superscriptaddress,longbibliography,floatfix]{revtex4-2}

\usepackage{amsmath,amssymb,bm}
\usepackage{graphicx}
\usepackage[normalem]{ulem}
\usepackage{hyperref}
\usepackage{booktabs}
\usepackage{orcidlink}
\hypersetup{colorlinks=true,citecolor=blue,urlcolor=blue,linkcolor=blue}
\usepackage{xcolor}

\newcommand{\kb}{k_{\mathrm B}}
\newcommand{\dd}{\mathrm d}
\newcommand{\Li}{\operatorname{Li}}
\newcommand{\zR}{\zeta_{\mathrm R}}
\newcommand{\G}{\mathcal{G}}
\newcommand{\e}{\mathrm e}
\begin{document}

\title{Coupling-Response Thermodynamic Density of States for Bosonic Working Media}

\author{Francisco J. Pe\~na\,\orcidlink{0000-0002-7432-0707}}
\affiliation{Facultad de Ingenier\'ia, Universidad San Sebasti\'an, Lago Panguipulli 1390, Puerto Montt, Chile}

\author{Bastian Castorene\,\orcidlink{0009-0002-9075-5716}}
\email{bastian.castorene.c@mail.pucv.cl}
\affiliation{Instituto de Física, Pontificia Universidad Católica de Valparaíso, Casilla 4950, 2373223 Valparaíso, Chile}
\affiliation{Departamento de Física, Universidad Técnica Federico Santa María, 2390123 Valparaíso, Chile}

\author{Nicol\'as Vidal-Silva \orcidlink{0000-0002-3717-3816}}
\affiliation{Departamento de Ciencias F\'isicas, Universidad de La Frontera, Casilla 54-D, Temuco, Chile}

\date{\today}


\begin{abstract}
We introduce an operational thermodynamic density of states (TDOS), $\G_\lambda=(\partial S/\partial\lambda)_{T,\mu}$, to characterize bosonic working media under microscopic spectral control. At fixed chemical potential, the response is strictly driven by the structural density of states, yielding the exact scaling law $\G_\lambda=-(d/\alpha)S/\lambda$ for homogeneous spectra $\epsilon(k)=\Delta+\lambda k^\alpha$, while coupling-dependent gaps generate edge terms only for finite boundary weight. Under a fixed-particle constraint, a secondary chemical-potential redistribution channel emerges. Exact solutions for quadratic bands and harmonic traps demonstrate that spatial dimensionality dictates whether this constrained TDOS exhibits a jump discontinuity or remains continuous across the Bose--Einstein condensation threshold. For pure scale control, the response maps exactly to the heat capacity, $\lambda \G_\lambda^{(N)}=-C_N$. This framework is applied to evaluate reversible work in Stirling cycles and to identify observables in magnonic platforms, contrasting global spectral edges with topological Dirac gaps in honeycomb ferromagnets.
\end{abstract}

\maketitle

\section{Introduction}

Thermodynamic response functions provide one of the most direct bridges between microscopic spectra and macroscopic behavior that measure how a physical system reacts to external changes in experimental setups. In electronic systems, the thermodynamic density of states (TDOS) is commonly associated with the isothermal charge compressibility $(\partial N/\partial\mu)_T$~\cite{AshcroftMermin,Datta,WOS:000755063500010,WOS:A1992JT05900021,WOS:A1993LW02600062}, because the particle number is conserved and the chemical potential ($\mu$) can be shifted by electrostatic gating, doping, or reservoir engineering \cite{WOS:000445327300008, WOS:000353867000068,WOS:A1995QE30500007,WOS:000267868700008}. This response governs Thomas--Fermi screening, quantum capacitance, and low-temperature carrier accumulation in nanoscale conductors \cite{Luryi1988,Eisenstein1992,Ponomarenko2008}. Its physical interpretation is especially transparent: at zero temperature, the thermal broadening window collapses and the thermodynamic derivative reduces to the structural density of states evaluated at the Fermi energy $(\epsilon_F)$.

In contrast, bosonic working media require a different thermodynamic viewpoint. Unlike electronic devices, where carrier accumulation is naturally controlled through the chemical potential, many bosonic platforms are operated through externally tunable Hamiltonian parameters that reshape the energy excitation spectrum. Examples include magnetic fields in spin systems, trap frequencies in ultracold gases~\cite{PethickSmith2008,PitaevskiiStringari2003}, cavity parameters in photonic platforms and superconducting circuits (cQED) \cite{Pekola2015, Campisi2015, Karimi2016, Cottet2017}, charging parameters in quantum batteries \cite{Alicki2013, Campaioli2017, Ferraro2018, Andolina2019}, and interaction or hopping amplitudes in lattice realizations \cite{Jaksch1998, Greiner2002, Bloch2008, Gross2017,WOS:000472681100013,WOS:001840550200009,WOS:001493202700001,10.1063/5.0011887,Cruz_2023}. Even when a bosonic chemical potential can be defined thermodynamically, spectral control often provides the experimentally relevant work coordinate. Consequently, the standard fermionic TDOS is not always the most natural descriptor of bosonic thermodynamic operation. This observation motivates a coupling-based reformulation of the TDOS concept for bosons, establishing a response function that quantifies how the thermally accessible spectral manifold changes under controlled deformations of the underlying Hamiltonian. This framework is directly relevant in the context of bosonic heat engines, where external operation is typically implemented not by modifying a particle reservoir, but by tuning Hamiltonian parameters \cite{Bauer2012,Chumak2015,Kosloff2014,Benenti2017}. Such parameters act as generalized work coordinates. Their conjugate forces are derived from the appropriate thermodynamic potential, and their temperature dependencies are linked to entropy sensitivities via Maxwell relations. Therefore, the central object of this work is the coupling-response TDOS,
\begin{equation}
\G_\lambda(T,\mu;\lambda)\equiv \left(\frac{\partial S}{\partial\lambda}\right)_{T,\mu},
\label{eq:def_G}
\end{equation}
{Physically, \(\mathcal G_\lambda\) measures the sensitivity of equilibrium
entropy to a controlled deformation of the Hamiltonian at fixed
temperature and chemical potential. A small parameter change produces
\(dS=\mathcal G_\lambda d\lambda\), so the sign of the response
distinguishes entropy enhancement from suppression upon increasing
\(\lambda\). In a reversible isothermal protocol, the associated heat
exchange is \(\delta Q_{\mathrm{rev}}=T\mathcal G_\lambda d\lambda\).
Equivalently, the Maxwell relation identifies \(\mathcal G_\lambda\)
with the temperature derivative of the conjugate generalized force.
These relations give the response an operational meaning, while its
units remain entropy per unit coupling rather than those of a
microscopic state-counting density. A protocol conserving particle
number \eqref{eq:G_fixedN_chain} instead probes \(\mathcal G_\lambda^{(N)}\), which includes the
chemical-potential adjustment required by that constraint.}

This terminology warrants clarification. The quantity introduced in Eq.~\eqref{eq:def_G} is not a structural density of states in the same sense as $g(\epsilon_F)$ for a zero-temperature Fermi liquid. It possesses the dimensions of entropy per unit coupling and is therefore not a density of states in the microscopic counting sense. We employ the TDOS terminology in an operational thermodynamic sense: $\G_\lambda$ is a DOS-resolved response that quantifies the sensitivity of the thermally populated spectral manifold to an externally controlled deformation. The novelty lies not in the formal existence of the derivative $(\partial S/\partial\lambda)_T$, which is a standard thermodynamic object, but rather in the identification of its spectral content, its fixed-$\mu$ versus fixed-$N$ decomposition, its edge dependence, and its fundamental role in coupling-controlled bosonic work cycles.
For instance, when bosons are driven by an external magnetic field $B$, identifying $\lambda=B$ reduces Eq.~\eqref{eq:def_G} to a generalized magnetocaloric Maxwell response. However, the present formulation is not restricted to magnetic fields. It extends this thermodynamic logic to arbitrary Hamiltonian couplings and resolves the associated spectral DOS content through $(\partial g/\partial\lambda)_\epsilon$. This spectral formulation enables the distinction between a purely DOS-driven fixed-$\mu$ channel and the supplementary redistribution channel that emerges under a particle-number constraint. Furthermore, it clarifies a crucial practical limitation: while the response $\G_\lambda$ quantifies the thermodynamic sensitivity per infinitesimal variation of a selected coordinate $\lambda$, assessing the true experimental advantages among different controls requires additionally accounting for accessible parameter ranges, control precision, dissipation, and the energetic cost of tuning the parameter.

In this manuscript, we develop this framework of the bosonic TDOS for noninteracting bosonic bands and use analytically solvable models to expose the structure of the response. At a fixed chemical potential, transforming from momentum space to energy space shows that all explicit dispersion sensitivity is absorbed into the parametric dependence of $g(\epsilon;\lambda)$ \cite{WOS:001767314800001}. The response is then purely DOS-driven. By contrast, if the total particle number is externally constrained, as in pumped bosonic gases approaching quasi-equilibrium steady states \cite{Demokritov2006,Bender2012,Rueckriegel2014,Bunkov2010,Bozhko2016,Divinskiy2021,Cornelissen2015,Cornelissen2016}, the chemical potential becomes an implicit function $\mu(T,\lambda)$. This generates a second contribution arising from the redistribution of occupations within the bosonic thermal window. We show that this implicit-$\mu$ channel cannot be eliminated by a simple reparametrization of the structural DOS, but requires a modification in the definition of the usual TDOS.

The remainder of the paper is organized as follows: Section~\ref{sec:fermion} reviews the fermionic TDOS and clarifies why the same compressibility-based construction is not the natural descriptor for coupling-controlled bosons. Section~\ref{sec:bosonic} defines the coupling-response TDOS, derives its Maxwell representation, and establishes the fixed-$\mu$ and fixed-$N$ decompositions. Section~\ref{sec:analytical_models} applies this framework to analytically solvable models, establishing the general scaling law $\G_\lambda=-(d/\alpha)S/\lambda$ for homogeneous spectra, incorporating boundary terms from coupling-dependent gaps, and solving fixed-$N$ protocols to clarify the dimensional origin of bounded versus threshold responses. Section~\ref{sec:HO} analyzes harmonic-trap protocols across the Bose--Einstein condensation threshold, demonstrating that the fixed-$N$ coupling-response TDOS exhibits a finite jump in a three-dimensional isotropic trap but remains continuous in the two-dimensional case. Section~\ref{sec:work} connects this response to isothermal work in Stirling-type cycles. Section~\ref{sec:magnons} links the framework to experimental observables in magnonic systems~\cite{Serga2010,Uchida2010} and applies it to the low-energy sectors of honeycomb ferromagnetic magnons~\cite{Kim2016,Owerre2016,Chen2018,Cai2021}, clarifying the distinction between true spectral thresholds and finite-energy Dirac gaps. Finally, our conclusions are presented in Sec.~\ref{sec:conclusion}.

\section{The conventional fermionic compressibility}
\label{sec:fermion}

In electronic systems, the conventional TDOS is defined through the isothermal particle-number response,
\begin{equation}
G_F(T,\mu)\equiv \left(\frac{\partial N}{\partial \mu}\right)_T,
\label{eq:fermionic_tdos}
\end{equation}
where $N$ denotes the total carrier number and $\mu$ is the chemical potential. Given a structural density of states $g(\epsilon)$ and the Fermi--Dirac distribution $f(\epsilon;T,\mu)=[\exp\{\beta(\epsilon-\mu)\}+1]^{-1}$, with $\beta=1/(k_BT)$, the total particle number is given by
\begin{equation}
N(T,\mu)=\int_{-\infty}^{\infty}g(\epsilon)f(\epsilon;T,\mu)\,\dd\epsilon.
\label{eq:N_fermion}
\end{equation}
Consequently, the TDOS can be expressed as
\begin{equation}
G_F(T,\mu)=\int_{-\infty}^{\infty}g(\epsilon)\left(-\frac{\partial f}{\partial\epsilon}\right)\dd\epsilon.
\label{eq:GF_window}
\end{equation}
The kernel $-\partial f/\partial\epsilon$ acts as a normalized thermal window centered at $\mu$. In the limit $T\to0$, it converges to $\delta(\epsilon-\epsilon_F)$, and therefore
\begin{equation}
G_F(0,\epsilon_F)=g(\epsilon_F).
\label{eq:TDOS_zeroT}
\end{equation}
At low but finite temperatures, the Sommerfeld expansion gives
\begin{equation}
G_F(T,\mu)=g(\mu)+\frac{\pi^2}{6}(\kb T)^2\left.\frac{\partial^2g}{\partial\epsilon^2}\right|_{\epsilon=\mu}+O(T^4),
\label{eq:Sommerfeld}
\end{equation}
provided that $g(\epsilon)$ is smooth on the thermal scale. This sequence of relations explains why Eq.~\eqref{eq:fermionic_tdos} is termed a TDOS: it is a thermodynamic response that reduces strictly to the structural DOS at the Fermi surface.

For bosonic working media, the compressibility-based definition is not always the most operational descriptor. Some bosonic systems possess a conserved or approximately conserved particle number and a well-defined chemical potential. However, many experimental realizations are controlled primarily through Hamiltonian parameters that modify the excitation spectrum. In such situations, the relevant thermodynamic question is not how the population changes under variations of $\mu$, but how the entropy responds to controlled spectral deformations.

For nonconserved bosonic excitations, such as magnons in thermal equilibrium, this distinction becomes even more pronounced. Their number is not conserved, and the equilibrium chemical potential is fixed by the statistical condition, commonly taking $\mu=0$ when excitation energies are measured relative to the magnetic ground state. In driven bosonic gases, an effective chemical potential may emerge, but Bose statistics constrain it to remain below the spectral threshold, $\mu\leq\Delta$. Therefore, while $(\partial N/\partial\mu)_T$ can still be defined in quasi-equilibrium situations, it is generally not the natural control response for equilibrium operation.

In many experimentally relevant situations, the physically accessible control parameters are Hamiltonian couplings. A magnetic field may shift spectral gaps, a trapping potential may modify the confinement energy scale, an interaction parameter may reshape the dispersion, and other external controls may continuously deform the excitation spectrum. These couplings constitute externally tunable work coordinates. If $\lambda$ denotes one such coordinate, then the relevant thermodynamic question is how the entropy responds when the spectrum is deformed at a fixed temperature. This leads naturally to the coupling-response TDOS defined in Eq.~\eqref{eq:def_G}. In this sense, the proposed bosonic construction is not a direct analogue of the fermionic compressibility. Rather, it provides an operational thermodynamic response that characterizes how the thermally accessible spectral manifold changes under deformations of the microscopic Hamiltonian, which will be analyzed in the next section.

{Thus, Table~\ref{tab:comparison} highlights the distinct thermodynamic roles
of the two response functions. The conventional fermionic TDOS quantifies
particle accumulation in response to a change in chemical potential, whereas
\(\mathcal G_\lambda\) quantifies the entropy response to the variation of a
Hamiltonian parameter. The bosonic particle-number susceptibility remains
well defined whenever the ensemble permits number fluctuations; the coupling
response is a complementary quantity that is particularly relevant when
spectral parameters serve as experimentally accessible work coordinates.

\begin{table*}[t]

\centering
\caption{Comparison of the responses in Eqs.~\eqref{eq:def_G} and \eqref{eq:fermionic_tdos} of the
manuscript. The fermionic DOS limit refers to the independent-particle
description used there.}
\label{tab:comparison}
\small
\setlength{\tabcolsep}{4pt}
\begin{tabular}{p{0.18\textwidth}p{0.36\textwidth}p{0.36\textwidth}}
\toprule
Property & Fermionic TDOS & Bosonic coupling response\\
\midrule

Definition
&
\(G_F=(\partial N/\partial\mu)_{T,\lambda}\)
&
\(\mathcal G_\lambda=(\partial S/\partial\lambda)_{T,\mu}\)
\\

Controlled variable
&
Chemical potential \(\mu\)
&
Hamiltonian parameter \(\lambda\)
\\

Response
&
Particle-number change
&
Entropy change
\\

Units
&
Inverse energy, with \(N\) a particle count
&
Entropy per unit coupling
\\

Sign
&
Nonnegative for a stable grand-canonical system
&
Either sign, depending on the spectral deformation
\\

Zero-temperature behavior
&
\(G_F\to g(\epsilon_F)\); this value can be zero
&
\(\mathcal G_\lambda\to0\) for regular gapped models at fixed
\(\Delta-\mu>0\)
\\

Typical context
&
Electrostatic gating, carrier accumulation, and quantum capacitance
&
Field, stiffness, trap-frequency, or other spectral-control protocols
\\

Operational connection
&
Population response to a chemical-potential change
&
Reversible isothermal heat and temperature dependence of the
conjugate force
\\

Constraint dependence
&
Requires specification of the remaining fixed controls
&
A fixed-\(N\) protocol instead probes \(\mathcal G_\lambda^{(N)}\)
\\

\bottomrule
\end{tabular}
\end{table*}}

\section{Bosonic coupling-response TDOS}
\label{sec:bosonic}

We consider $d$-dimensional noninteracting bosonic bands $\{\epsilon_\nu(\mathbf{k};\lambda)\}$ controlled by a parameter $\lambda$. The grand potential is given by
\begin{equation}
\Omega(T,\mu;\lambda)=\kb T\sum_\nu\int\frac{\dd^d k}{(2\pi)^d}\ln\left[1-\e^{-\beta(\epsilon_\nu(\mathbf{k};\lambda)-\mu)}\right].
\label{eq:Omega_k}
\end{equation}
Equivalently, by introducing the structural density of states,
\begin{equation}
g(\epsilon;\lambda)=\sum_\nu\int\frac{\dd^d k}{(2\pi)^d}\delta[\epsilon-\epsilon_\nu(\mathbf{k};\lambda)],
\label{eq:DOS_def}
\end{equation}
the grand potential can be expressed as
\begin{equation}
\Omega(T,\mu;\lambda)=\kb T\int\dd\epsilon\,g(\epsilon;\lambda)\ln\left[1-\e^{-\beta(\epsilon-\mu)}\right].
\label{eq:Omega_energy}
\end{equation}
The corresponding entropy takes the form
\begin{equation}
S(T,\mu;\lambda)=\kb\int\dd\epsilon\,g(\epsilon;\lambda)s_B(\epsilon;T,\mu),
\label{eq:S_DOS}
\end{equation}
where $s_B$ represents the dimensionless entropy per bosonic mode,
\begin{equation}
s_B=(1+n_B)\ln(1+n_B)-n_B\ln n_B,
\label{eq:sB}
\end{equation}
with the Bose--Einstein distribution defined as
\begin{equation}
n_B(\epsilon;T,\mu)=\frac{1}{\e^{\beta(\epsilon-\mu)}-1}.
\label{eq:nB}
\end{equation}
The generalized force conjugate to $\lambda$ is defined as
\begin{equation}
X_\lambda=-\left(\frac{\partial\Omega}{\partial\lambda}\right)_{T,\mu}.
\label{eq:X_def}
\end{equation}

Since $S=-(\partial\Omega/\partial T)_{\mu,\lambda}$, equality of mixed derivatives gives the Maxwell relation
\begin{equation}
\G_\lambda^{\text{DOS}}(T,\mu;\lambda)=\left(\frac{\partial S}{\partial\lambda}\right)_{T,\mu}=\left(\frac{\partial X_\lambda}{\partial T}\right)_{\lambda,\mu}.
\label{eq:Maxwell}
\end{equation}
This identity makes the response experimentally meaningful: it may be obtained either from coupling-dependent entropy reconstruction or from the temperature dependence of the conjugate generalized force.

Differentiating Eq.~\eqref{eq:S_DOS} with respect to $\lambda$ at fixed $T$ and $\mu$ yields
\begin{equation}
\G_\lambda^{\text{DOS}}=\kb\int\dd\epsilon\left[\left(\frac{\partial g(\epsilon;\lambda)}{\partial\lambda}\right)_\epsilon s_B(\epsilon;T,\mu)+g(\epsilon;\lambda)\left(\frac{\partial s_B}{\partial\lambda}\right)_{T,\mu,\epsilon}\right].
\label{eq:G_general}
\end{equation}
In the energy representation, $s_B(\epsilon;T,\mu)$ depends on $\lambda$ only through $T$, $\mu$, or the integration variable $\epsilon$. Consequently, at fixed $(T,\mu,\epsilon)$, the second term vanishes, reducing the expression to
\begin{equation}
\G_\lambda^{\text{DOS}}=\kb\int\dd\epsilon\left(\frac{\partial g(\epsilon;\lambda)}{\partial\lambda}\right)_\epsilon s_B(\epsilon;T,\mu).
\label{eq:G_fixedmu}
\end{equation}
Eq.~\eqref{eq:G_fixedmu} constitutes the central fixed-$\mu$ result. It does not imply that the microscopic dispersion is irrelevant. Instead, it establishes that once the problem is expressed in the energy representation, all dependence on the deformation of $\epsilon_\nu(\mathbf{k};\lambda)$ is completely encapsulated by the induced density of states $g(\epsilon;\lambda)$. Therefore, the apparent momentum-space term proportional to $(\partial s_B/\partial\epsilon)(\partial\epsilon/\partial\lambda)$ is not an independent contribution in the spectral representation, as it is intrinsically absorbed into the derivative of the DOS.

Under a fixed particle-number constraint, the chemical potential $\mu$ is no longer an independent variable. Instead, the condition
\begin{equation}
N(T,\mu(T;\lambda);\lambda)=\int\dd\epsilon\,g(\epsilon;\lambda)n_B(\epsilon;T,\mu(T;\lambda))=N_0
\label{eq:N_fixed}
\end{equation}
defines $\mu$ as an implicit parametric function $\mu(T;\lambda)$. Consequently, the fixed-$N$ response is obtained by applying the chain rule to the entropy $S(T,\mu(T;\lambda);\lambda)$, yielding
\begin{equation}
\G^{(N)}_\lambda\equiv\left(\frac{\partial S}{\partial\lambda}\right)_{T,N}=\left(\frac{\partial S}{\partial\lambda}\right)_{T,\mu}+\left(\frac{\partial S}{\partial\mu}\right)_{T,\lambda}\left(\frac{\partial\mu}{\partial\lambda}\right)_{T,N}.
\label{eq:G_fixedN_chain}
\end{equation}
The required derivative of the chemical potential is determined by differentiating the constraint $N=N_0$ at constant $T$, via the implicit-function theorem:
\begin{equation}
\left(\frac{\partial\mu}{\partial\lambda}\right)_{T,N}=-\frac{(\partial N/\partial\lambda)_{T,\mu}}{(\partial N/\partial\mu)_{T,\lambda}}.
\label{eq:dmu_dlambda}
\end{equation}
and thus,
\begin{equation}
\G^{(N)}_\lambda=\G^{\mathrm{DOS}}_\lambda-\left(\frac{\partial S}{\partial\mu}\right)_{T,\lambda}\frac{(\partial N/\partial\lambda)_{T,\mu}}{(\partial N/\partial\mu)_{T,\lambda}}.
\label{eq:G_fixedN_decomp}
\end{equation}
This expression separates the fixed-$\mu$ DOS channel from a redistribution channel induced by the constraint. The latter is intrinsically thermodynamic: it arises because maintaining $N=N_0$ requires a compensating shift in the occupation function when $\lambda$ changes.

\section{Analytical models of the bosonic coupling response}
\label{sec:analytical_models}

To illustrate the formal framework developed in the preceding sections, we now evaluate the coupling-response TDOS for several paradigmatic bosonic spectra. We analyze these systems progressively, starting from fundamental scaling behaviors, incorporating the effects of moving spectral edges, and finally examining the consequences of a strict particle-number constraint.

\subsection{Homogeneous gapped spectra and scaling laws}
\label{subsec:homogeneous}
 We first consider a generic single isotropic gapped band
\begin{equation}
\epsilon(k;\lambda)=\Delta+\lambda k^\alpha,\qquad \Delta>0,\quad \lambda>0,\quad \alpha>0,
\label{eq:homogeneous_disp}
\end{equation}
with fixed gap and fixed chemical potential. The density of states has the scaling form
\begin{equation}
g_d(\epsilon;\lambda)=C_{d,\alpha}\lambda^{-d/\alpha}(\epsilon-\Delta)^{d/\alpha-1}\Theta(\epsilon-\Delta),
\label{eq:homogeneous_DOS}
\end{equation}
where $d$ denotes the spatial dimensionality, $C_{d,\alpha}$ is a geometry-dependent constant and $\Theta$ is the Heaviside step-function. Since the full $\lambda$ dependence is the prefactor $\lambda^{-d/\alpha}$, Eq.~\eqref{eq:G_fixedmu} immediately gives
\begin{equation}
\G^{\mathrm{DOS}}_\lambda=-\frac{d}{\alpha}\frac{S}{\lambda}.
\label{eq:scaling_general}
\end{equation}
This is the general homogeneous-band scaling law. { Thus, the coefficient \(d/\alpha\) follows from
the scaling of momentum-space volume and is independent of the explicit
evaluation of the thermal integrals. The relation is exact within this
spectral class.} For quadratic spectra ($\alpha=2$), it reduces to $\G_\lambda=-(d/2)S/\lambda$, whereas for a one-dimensional linear spectrum ($d=1, \alpha=1$), it yields $\G_\lambda=-S/\lambda$. This distinction prevents the quadratic scaling law from being incorrectly applied to a linear benchmark.

To illustrate this, we evaluate the thermodynamic functions explicitly by introducing the fugacity $z=\exp[\beta(\mu-\Delta)]$. The 1D linear and 2D quadratic bands share a density of states scaling as $\lambda^{-1}$, yielding $g(\epsilon;\lambda) = \Gamma \lambda^{-1}\Theta(\epsilon-\Delta)$, with geometric factors $\Gamma_{1\mathrm{D}}=L/\pi$ and $\Gamma_{2\mathrm{D}}=A/(4\pi)$. Their shared scaling produces functionally identical grand potentials and entropies:
\begin{align}
\Omega &= -\frac{\Gamma}{\lambda}(\kb T)^2 \Li_2(z), \\ S &= \frac{\Gamma\kb}{\lambda}\left[(\Delta-\mu)\Li_1(z)+2 \kb T \Li_2(z)\right].
\label{eq:thermo_1D_2D}
\end{align}
where $\Li_n(z) = \sum_{k=1}^\infty z^k/k^n$ denotes the polylogarithm function. Consequently, their generalized forces and coupling-response TDOS strictly satisfy $X_\lambda = \Omega/\lambda$ and $\G^{\mathrm{DOS}}_\lambda = -S/\lambda$, respectively, as required by $d/\alpha=1$.

Conversely, the 3D quadratic band ($d=3, \alpha=2$) scales as $\lambda^{-3/2}$, with a structural density of states
\begin{equation}
g_{3\mathrm{D}}(\epsilon;\lambda)=\frac{V}{4\pi^2\lambda^{3/2}}\sqrt{\epsilon-\Delta}\,\Theta(\epsilon-\Delta).
\label{eq:DOS_3D}
\end{equation}
The corresponding grand potential and entropy are
\begin{equation}
\Omega_{3\mathrm{D}}=-\frac{V}{4\pi^2\lambda^{3/2}}(\kb T)^{5/2}\frac{\sqrt{\pi}}{2}\Li_{5/2}(z),
\label{eq:Omega_3D}
\end{equation}
\begin{equation}
S_{3\mathrm{D}}=\frac{V\kb}{4\pi^2\lambda^{3/2}}(\kb T)^{3/2}\frac{\sqrt{\pi}}{2}\left[\frac{\Delta-\mu}{\kb T}\Li_{3/2}(z)+\frac{5}{2}\Li_{5/2}(z)\right].
\label{eq:S_3D}
\end{equation}
The conjugate force and TDOS response then follow as $X_{\lambda,3\mathrm{D}} = \frac{3}{2}\Omega_{3\mathrm{D}}/\lambda$ and $\G^{\mathrm{DOS}}_{\lambda,3\mathrm{D}} = -\frac{3}{2}S_{3\mathrm{D}}/\lambda$.

These analytical examples confirm the organizing principle: the geometric prefactor dictating how the density of states scales with $\lambda$ strictly determines the macroscopic entropy response.
\subsection{Coupling-dependent gaps and spectral-edge terms}
\label{subsec:edge}

When a control parameter directly modifies the spectral energy threshold, the thermodynamic response acquires a boundary contribution. Consider the two-dimensional quadratic band with a coupling-dependent gap,
\begin{equation}
\epsilon(k;\lambda)=\Delta(\lambda)+\lambda k^2,\qquad \Delta(\lambda)>0.
\label{eq:gap_disp}
\end{equation}
The structural density of states is
\begin{equation}
g_{2\mathrm{D}}(\epsilon;\lambda)=\frac{A}{4\pi\lambda}\Theta[\epsilon-\Delta(\lambda)].
\label{eq:DOS_gap}
\end{equation}
Differentiating with respect to $\lambda$ yields both the bulk scaling term and a localized term at the moving threshold,
\begin{equation}
\left(\frac{\partial g_{2\mathrm{D}}}{\partial\lambda}\right)_\epsilon=-\frac{A}{4\pi\lambda^2}\Theta[\epsilon-\Delta(\lambda)]-\frac{A\Delta'(\lambda)}{4\pi\lambda}\delta[\epsilon-\Delta(\lambda)],
\label{eq:DOS_gap_deriv}
\end{equation}
where $\Delta'(\lambda)=\dd\Delta/\dd\lambda$. Substituting this into Eq.~\eqref{eq:G_fixedmu} separates the TDOS into bulk and edge responses:
\begin{equation}
\G_\lambda(T;\lambda)=-\frac{S(T;\lambda)}{\lambda}+\G^{\mathrm{edge}}_\lambda(T;\lambda),
\label{eq:G_edge_decomp}
\end{equation}
with
\begin{equation}
\G^{\mathrm{edge}}_\lambda(T;\lambda)=-\frac{A\kb}{4\pi}\frac{\Delta'(\lambda)}{\lambda}s_B[\Delta(\lambda);T,\mu].
\label{eq:G_edge}
\end{equation}
Physically, a gap that increases with $\lambda$ ($\Delta'>0$) gives a negative edge contribution because spectral weight is pushed to higher energies, depleting thermally accessible states. Conversely, a decreasing gap enhances the entropic response.

Equation~\eqref{eq:G_edge} applies strictly to a moving threshold whose structural DOS has a finite boundary weight. More generally, if the low-energy DOS takes the form
\begin{equation}
g(\epsilon;\lambda)=C(\lambda)[\epsilon-\Delta(\lambda)]^\eta\Theta[\epsilon-\Delta(\lambda)],
\label{eq:DOS_edge_general}
\end{equation}
differentiating the threshold produces a strictly localized delta-function term only if the edge exponent $\eta=0$. For $\eta>0$, the factor $[\epsilon-\Delta(\lambda)]^\eta$ suppresses the boundary singularity, and the gap shift instead modifies the integrated thermal population above the edge (while $\eta<0$ requires specific physical regularization). This threshold-exponent dependence is crucial for the honeycomb-magnon application discussed below, where a zone-center quadratic edge and a finite-energy Dirac gap exhibit fundamentally different thermodynamic behaviors.

\subsection{Fixed-particle protocols and the implicit chemical-potential channel}
\label{subsec:fixedN}

While the previous examples isolated the density-of-states contribution at a fixed chemical potential, constrained experimental setups activate a secondary redistribution channel. We explicitly solve fixed-$N$ protocols, which are relevant for pumped bosonic gases where the population can be approximately stabilized over the relaxation time of the quasi-equilibrium state \cite{Demokritov2006,Bender2012,Rueckriegel2014,Bozhko2016,Divinskiy2021}. 

For the two-dimensional quadratic band $\epsilon=\Delta+\lambda k^2$, imposing $N=N_0$ yields the chemical potential $\mu(T,\lambda)=\Delta+\kb T\ln(1-\e^{-\chi})$, where $\chi=4\pi\lambda N_0/(A\kb T)$ is the dimensionless density. The constrained derivative,
\begin{equation}
\left(\frac{\partial\mu}{\partial\lambda}\right)_{T,N}=\frac{4\pi N_0}{A(\e^\chi-1)},
\label{eq:dmu_dlambda_2D}
\end{equation}
is strictly positive: increasing $\lambda$ reduces the DOS prefactor, forcing $\mu$ upward toward the band edge to conserve $N$.

\begin{figure*}[t]
\centering
\begin{minipage}[t]{0.48\textwidth}
\centering
\includegraphics[width=\linewidth]{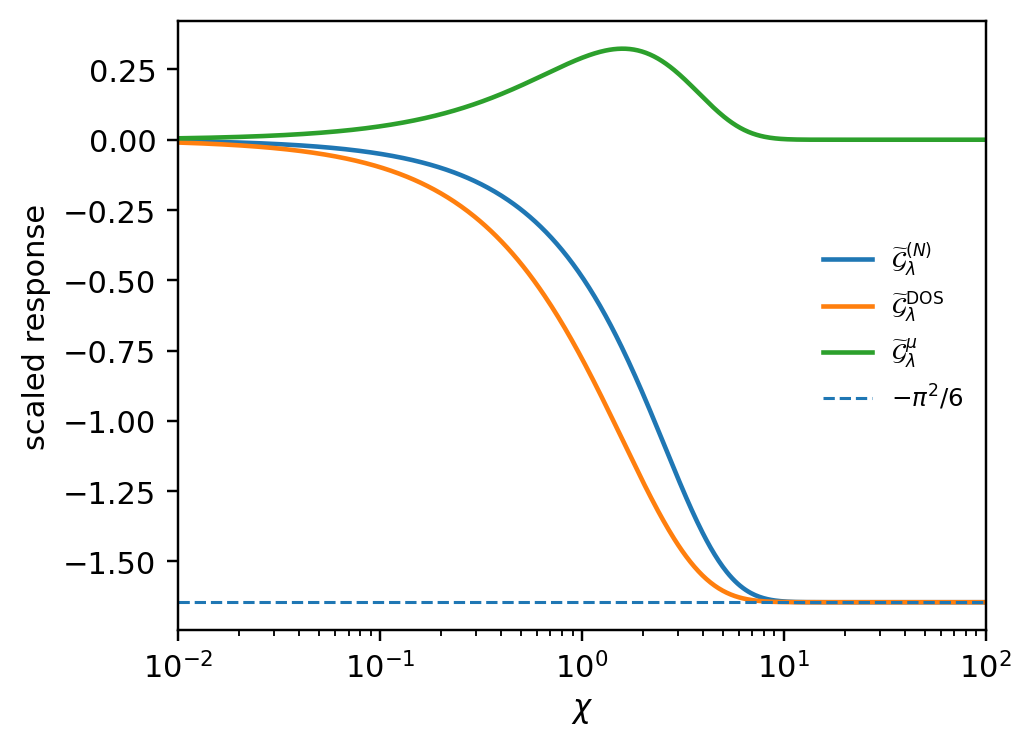}\par
\vspace{0.2em}
(a)
\end{minipage}
\hfill
\begin{minipage}[t]{0.48\textwidth}
\centering
\includegraphics[width=\linewidth]{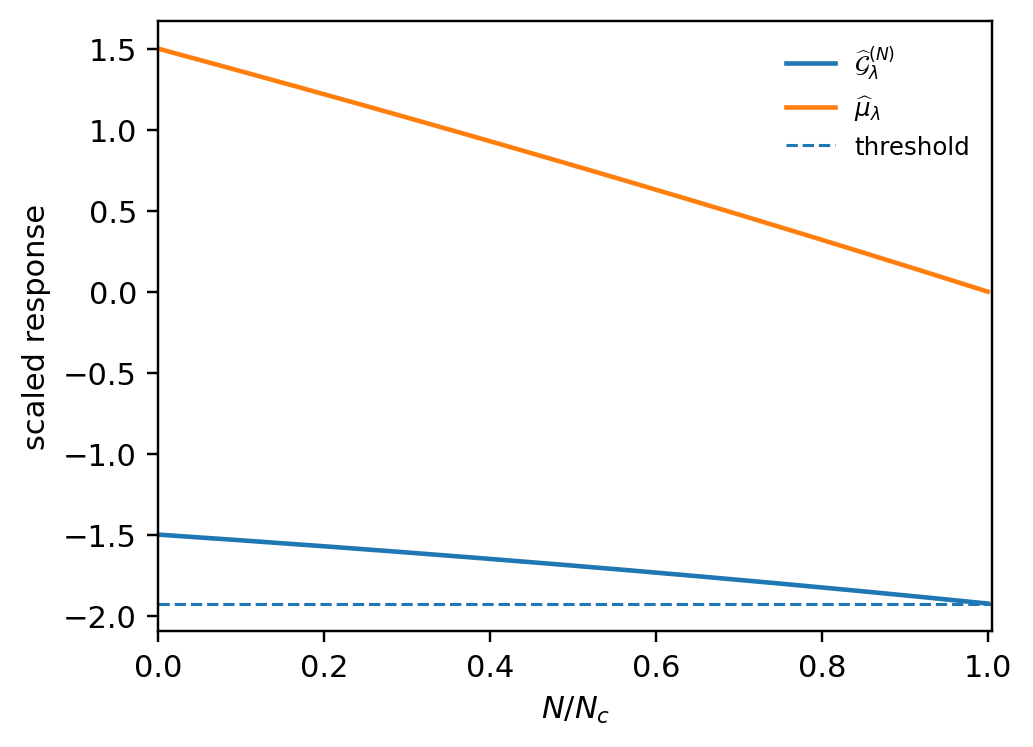}\par
\vspace{0.2em}
(b)
\end{minipage}
\caption{Fixed-particle coupling-response TDOS for quadratic bosonic bands. (a) Dimensionless two-dimensional response, scaled as $\widetilde{\G}_\lambda=(2\pi\lambda^2/A\kb^2T)\G_\lambda$, decomposed into the DOS-driven channel $\widetilde{\G}^{\mathrm{DOS}}_\lambda$ and the implicit-chemical-potential channel $\widetilde{\G}^{\mu}_\lambda$. The parameter $\chi=4\pi\lambda N_0/(A\kb T)$ controls the crossover from the dilute regime to the degenerate two-dimensional regime; the positive $\mu$-induced contribution partially compensates the negative DOS-driven response and vanishes as $\mu\to\Delta^-$, leaving $\widetilde{\G}^{(N)}_\lambda\to-\pi^2/6$. (b) Three-dimensional normal-phase response as a function of the reduced population $N/N_c$. The scaled response is $\widehat{\G}^{(N)}_\lambda=(\lambda/N\kb)\G^{(N)}_\lambda$, while $\widehat{\mu}_\lambda=(\lambda/\kb T)(\partial\mu/\partial\lambda)_{T,N}$ measures the implicit chemical-potential channel. As $N/N_c\to1^-$, the chemical potential becomes pinned to the spectral edge, $\widehat{\mu}_\lambda\to0$, and the response approaches the finite threshold value $-(15/4)\zR(5/2)/\zR(3/2)$.}
\label{fig:fixedN}
\end{figure*}

Substituting $\mu(T,\lambda)$ into the entropy and differentiating allows us to express the dimensionless response, $\widetilde{\G}^{(N)}_\lambda = (2\pi\lambda^2/A\kb^2T)\G^{(N)}_\lambda$, as the exact sum of a DOS-driven channel and an implicit-$\mu$ redistribution channel:
\begin{align}
\begin{aligned}
    \widetilde{\G}^{(N)}_\lambda = \overbrace{\frac{\chi}{2}\ln(1-\e^{-\chi})-\Li_2(1-\e^{-\chi})}^{\widetilde{\G}^{\mathrm{DOS}}_\lambda} +\\ \underbrace{\frac{\chi^2}{2(\e^\chi-1)}-\frac{\chi}{2}\ln(1-\e^{-\chi})}_{\widetilde{\G}^{\mu}_\lambda}.
\end{aligned}
\label{eq:Gtilde_decomp_2D}
\end{align}
As illustrated in Fig.~\ref{fig:fixedN}(a), the negative DOS-driven response is partially compensated by the positive $\mu$-induced shift. In the classical dilute limit ($\chi\ll 1$), the total response scales as $\G^{(N)}_\lambda \simeq -N_0\kb/\lambda$. In the degenerate limit ($\chi\gg 1$, $\mu\to\Delta^-$), the implicit-$\mu$ channel is exponentially suppressed, and the response saturates to a finite value, $\G^{(N)}_\lambda \simeq -A\pi\kb^2T/(12\lambda^2)$.

The dimensional origin of this saturation lies in the edge DOS. Because the one-dimensional linear and two-dimensional quadratic bands have constant or diverging DOS at the threshold, their excited-state capacity grows logarithmically, preventing a finite-density critical singularity at nonzero temperature. Conversely, the three-dimensional quadratic band possesses a vanishing edge DOS, $g_{3\mathrm{D}}(\epsilon)\propto\sqrt{\epsilon-\Delta}$, which yields a finite critical capacity at $\mu=\Delta$:
\begin{equation}
N_c(T)=\frac{V}{4\pi^2\lambda^{3/2}}(\kb T)^{3/2}\frac{\sqrt{\pi}}{2}\zR(3/2),
\label{eq:Nc_3D}
\end{equation}
where $\zR(s)$ denotes the Riemann zeta function. Defining $z=\exp[\beta(\mu-\Delta)]$, the normal-phase population is $N/N_c=\Li_{3/2}(z)/\zR(3/2)$. Near the threshold ($z\to 1^-$), the divergence of the isothermal compressibility $(\partial N/\partial\mu)_{T,\lambda}$ pins the chemical potential, suppressing its constrained derivative $\widehat{\mu}_\lambda\equiv (\lambda/\kb T)(\partial\mu/\partial\lambda)_{T,N} = \frac{3}{2}\Li_{3/2}(z)/\Li_{1/2}(z)$. The resulting dimensionless response is
\begin{align}
\widehat{\G}^{(N)}_\lambda&\equiv\frac{\lambda}{N\kb}\G^{(N)}_\lambda, \nonumber \\
&=\frac{3}{2}\frac{\Li_{3/2}(z)}{\Li_{1/2}(z)}\left[\frac{5}{2}\left(1-\frac{\Li_{5/2}(z)\Li_{1/2}(z)}{\Li_{3/2}^2(z)}\right)-1\right].
\label{eq:Ghat_3D}
\end{align}
This response transitions from the dilute value $-3/2$ to the finite threshold value $-(15/4)\zR(5/2)/\zR(3/2)$ as $z\to 1^-$ shown in Fig.~\ref{fig:fixedN}(b) . The three-dimensional case is therefore distinguished by a genuine condensation threshold and an associated nonanalytic constrained crossover, rather than an automatic divergence of the total thermodynamic response \cite{Bunkov2010,Bozhko2016}.

\section{Harmonic-trap protocols across the BEC threshold}
\label{sec:HO}

The homogeneous benchmarks illustrate the response for power-law spectral edges. We now turn to confined systems that support a genuine finite-temperature Bose--Einstein condensate, specifically comparing the three- and two-dimensional isotropic harmonic traps.

\subsection{Three-dimensional isotropic harmonic trap}
\label{subsec:HO3D}

For a 3D isotropic trap with coupling $\lambda\equiv\hbar\omega$, the semiclassical density of states is $g_{\mathrm{HO},3}(\epsilon;\lambda)=\epsilon^2/(2\lambda^3)\Theta(\epsilon)$. This scales as $\lambda^{-3}$, yielding an effective dimension $d_{\mathrm{eff}}/\alpha=3$. Setting $\Delta=0$ and $z=\e^{\beta\mu}$, the particle number and entropy are
\begin{align}
N_{\mathrm{HO},3}&=\frac{(\kb T)^3}{\lambda^3}\Li_3(z),\\
S_{\mathrm{HO},3}&=\frac{(\kb T)^3\kb}{\lambda^3}[4\Li_4(z)-\beta\mu\Li_3(z)].
\label{eq:thermo_HO3}
\end{align}
Consequently, the fixed-$\mu$ response strictly follows the scaling law $\G_{\lambda,\mathrm{HO},3}^{\mathrm{DOS}}=-3S_{\mathrm{HO},3}/\lambda$. The condensation threshold, $\kb T_c(\lambda)=\lambda[N_0/\zR(3)]^{1/3}$, scales linearly with $\lambda$. 

In the normal phase ($T>T_c$), substituting the thermodynamic derivatives into Eq.~\eqref{eq:G_fixedN_decomp} explicitly yields the implicit-$\mu$ channel:
\begin{equation}
\G_\lambda^\mu=\frac{3N_0\kb}{\lambda}\left[3\frac{\Li_3(z)}{\Li_2(z)}-\beta\mu\right].
\label{eq:Gmu_HO3_explicit}
\end{equation}
In the dilute limit ($z\to 0$), the logarithmic divergences in the DOS and implicit-$\mu$ channels cancel, collapsing the total response to $\G_\lambda^{(N)}\simeq -3N_0\kb/\lambda$.

In the condensed phase ($T<T_c$), Bose statistics pin $\mu=0$. The implicit-$\mu$ channel vanishes identically, and the total fixed-$N$ response equals the purely thermal DOS channel: $\G_\lambda^{(N)}=-12\zR(4)\kb(\kb T)^3/\lambda^4$.

\begin{figure}[t]
\centering
\includegraphics[width=\columnwidth]{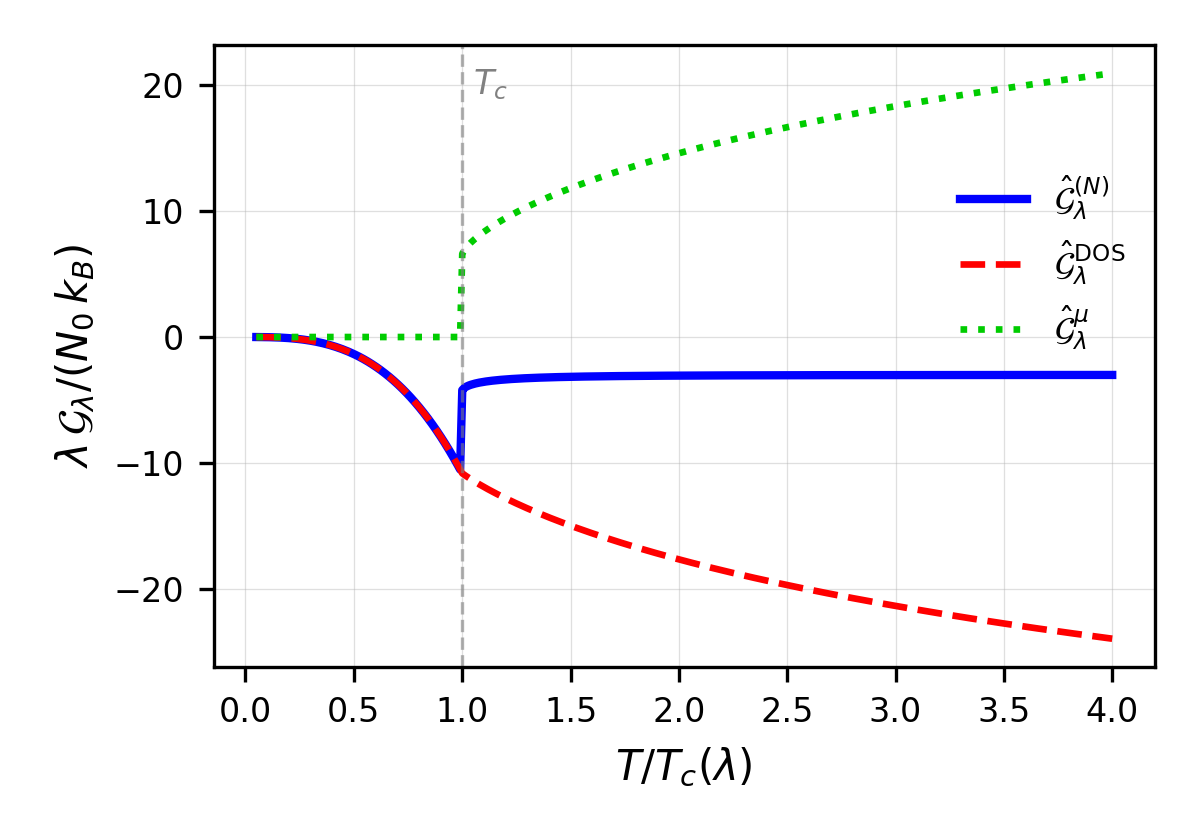}
\caption{{Dimensionless fixed-particle coupling-response TDOS for a three-dimensional isotropic harmonic trap, with $\lambda=\hbar\omega$ and $\widehat{\G}^{(N)}_\lambda=\lambda\G_\lambda^{(N)}/(N_0\kb)$. Solid blue: total. Dashed red: DOS channel. Dotted green: implicit-$\mu$ channel. Below $T_c$ the implicit-$\mu$ channel vanishes because $\mu$ is pinned to the band edge, and the total response equals the DOS channel. Above $T_c$ the two channels partially cancel. The jump at $T_c$ has the closed-form magnitude given in Eq.~\eqref{eq:G_HO3_jump}.}}
\label{fig:HO3D_TDOS}
\end{figure}

At $T=T_c(\lambda)$, approaching from above, $\Li_2(z)\to \pi^2/6$. Thus, the constrained derivative $(\partial\mu/\partial\lambda)_{T,N}$ remains finite, causing the implicit-$\mu$ channel (and thereby the total response) to exhibit a finite jump discontinuity:
\begin{equation}
\Delta\G_\lambda^{(N)}\equiv\G_\lambda^\mu(T_c^+)=\frac{54\zR(3)}{\pi^2}\frac{N_0\kb}{\lambda}.
\label{eq:G_HO3_jump}
\end{equation}
This behavior, illustrated in Fig.~\ref{fig:HO3D_TDOS}, contrasts sharply with the homogeneous 3D quadratic band, where a diverging compressibility forces a continuous response.

\subsection{Two-dimensional isotropic harmonic trap}
\label{subsec:HO2D}

The 2D trap exhibits a finite-temperature condensation but a qualitatively different constrained response. With $g_{\mathrm{HO},2}(\epsilon;\lambda)=\epsilon/\lambda^2\Theta(\epsilon)$, the effective dimension is $d_{\mathrm{eff}}/\alpha=2$. The corresponding population and entropy are
\begin{align}
N_{\mathrm{HO},2}&=\frac{(\kb T)^2}{\lambda^2}\Li_2(z), \nonumber \\ S_{\mathrm{HO},2}&=\frac{(\kb T)^2\kb}{\lambda^2}[3\Li_3(z)-\beta\mu\Li_2(z)],
\label{eq:thermo_HO2}
\end{align}
with a critical temperature $\kb T_c(\lambda)=\lambda\sqrt{6N_0}/\pi$. The DOS channel is $\G_{\lambda,\mathrm{HO},2}^{\mathrm{DOS}}=-2S_{\mathrm{HO},2}/\lambda$.

\begin{figure}[t]
\centering
\includegraphics[width=\columnwidth]{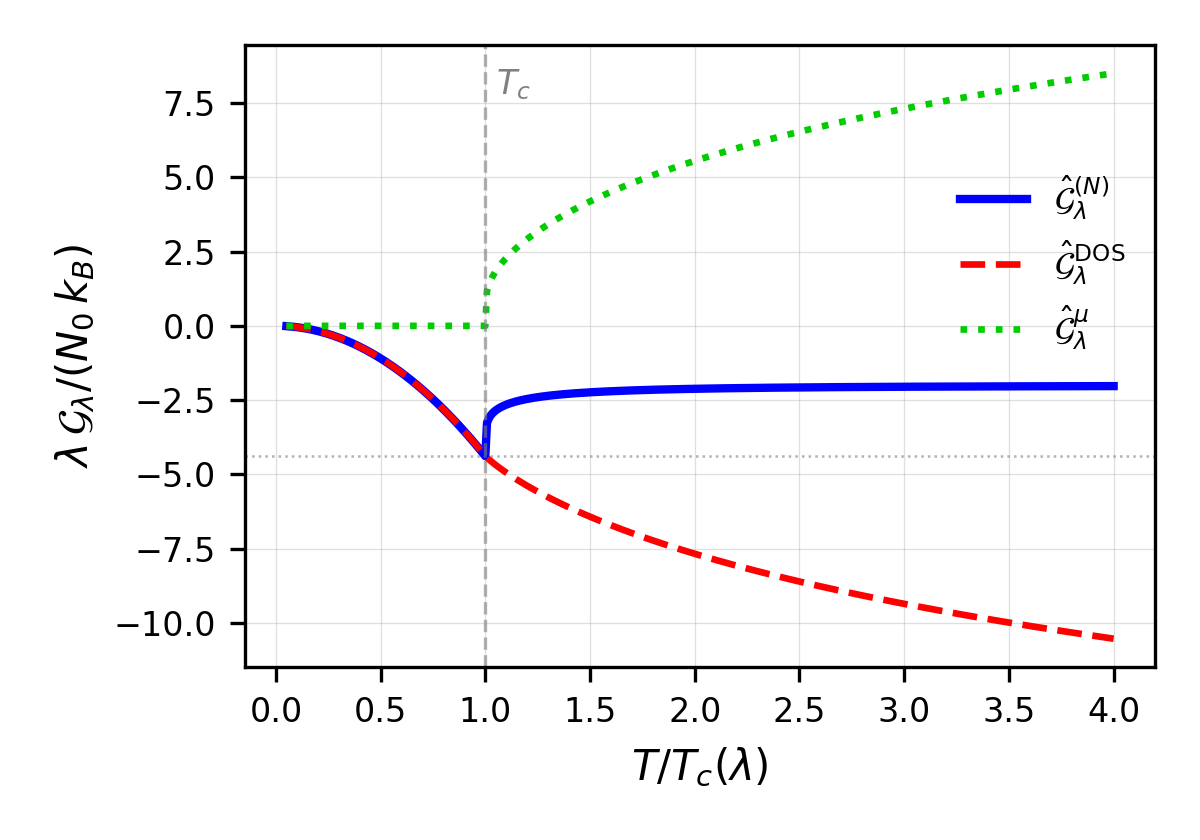}
\caption{{Dimensionless fixed-particle coupling-response TDOS for a two-dimensional isotropic harmonic trap. In contrast with the three-dimensional case (Fig.~\ref{fig:HO3D_TDOS}), the total response is continuous across $T_c$, taking the value $-36\zR(3)/\pi^{2}\simeq-4.385$. The implicit-$\mu$ channel vanishes continuously as $T\to T_c^{+}$ because $\Li_{1}(1)$ diverges in the denominator of $(\partial\mu/\partial\lambda)_{T,N}$, in sharp contrast with the finite limit of $\Li_{2}(1)$ that produces a finite jump in three dimensions. The vanishing is logarithmically slow, $\G^{\mu}_\lambda\propto1/|\ln(1-z)|$ in Eq.~\eqref{eq:Gmu_HO2}, so the total response rises steeply but continuously at $T_c^{+}$.}}
\label{fig:HO2D_TDOS}
\end{figure}

In the normal phase, the implicit-$\mu$ channel takes the form
\begin{equation}
\G_{\lambda,\mathrm{HO},2}^\mu=\frac{2N_0\kb}{\lambda}\left[2\frac{\Li_2(z)}{\Li_1(z)}-\beta\mu\right].
\label{eq:Gmu_HO2}
\end{equation}
Crucially, as $T\to T_c^+$ ($z\to 1^-$), the denominator $\Li_1(z)=-\ln(1-z)$ diverges logarithmically. Consequently, the implicit-$\mu$ channel vanishes continuously ($\G_\lambda^\mu\propto 1/|\ln(1-z)|\to 0$), yielding a continuous total fixed-$N$ response at $T_c$:
\begin{equation}
\G_{\lambda,\mathrm{HO},2}^{(N)}(T_c)=-\frac{36\zR(3)}{\pi^2}\frac{N_0\kb}{\lambda}.
\label{eq:GN_HO2_atTc}
\end{equation}
As shown in Fig.~\ref{fig:HO2D_TDOS}, the response merely changes slope rather than jumping.

The contrast between the two traps demonstrates that the dimensionality of the spectral edge fundamentally dictates whether the coupling-response TDOS jumps or remains continuous at $T_c$.

Physically, this distinction is intimately linked to the heat capacity. Because the fixed-$N$ constraint for both traps fixes the fugacity as a function of the single scale variable $\kb T/\lambda$, the constrained entropy satisfies $S=S(\kb T/\lambda)$. This yields the exact identity
\begin{equation}
\lambda\G_\lambda^{(N)}=-T\left(\frac{\partial S}{\partial T}\right)_{\lambda,N}=-C_N.
\label{eq:G_CN_identity}
\end{equation}
Therefore, the TDOS jump in Eq.~\eqref{eq:G_HO3_jump} perfectly mirrors the well-known specific-heat discontinuity of the ideal 3D Bose gas \cite{PethickSmith2008}, and the continuity in 2D reflects the continuity of its respective $C_N$. This identity holds whenever the control parameter acts as a pure scale factor, offering a direct calorimetric route to measure the coupling-response TDOS.

\section{Isothermal work and Stirling-type cycles}
\label{sec:work}

The coupling-response TDOS intrinsically governs the isothermal work extracted during a parameter stroke. Following the convention that $W_{\mathrm{iso}}>0$ denotes work performed by the bosonic working medium, the infinitesimal reversible work is $\delta W_{\mathrm{iso}}=X_\lambda \, \dd\lambda$. In a closed Stirling-type cycle, the isothermal branches contribute work via this generalized force, while the Maxwell relation in Eq.~\eqref{eq:Maxwell} dictates how $\G_\lambda$ controls the corresponding heat exchange when the coupling is modified.

For the fixed-$N$ two-dimensional quadratic band, the relevant thermodynamic potential is the Helmholtz free energy $F=\Omega+\mu N$. The generalized force is $X^{(N)}_\lambda=-(\partial F/\partial\lambda)_{T,N}$. Expressed in terms of the dimensionless density $\chi=4\pi\lambda N_0/(A\kb T)$, the finite isothermal work from $\lambda_i$ to $\lambda_f$ evaluates to
\begin{equation}
W^{(N)}_{\mathrm{iso}}=-N_0\kb T\int_{\chi_i}^{\chi_f}\frac{\Li_2(1-\e^{-\chi})}{\chi^2} \, \dd\chi.
\label{eq:W_chi}
\end{equation}

This integral reveals a fundamental crossover in the work output. Consider an idealized Stirling cycle operating between temperatures $T_H$ and $T_L$, where the hot isothermal branch is traversed from $\lambda_>$ to $\lambda_<$ and the cold branch in reverse. In the classical dilute limit ($\chi\ll 1$), $\Li_2(1-\e^{-\chi})\simeq\chi$, yielding a logarithmic work law characteristic of phase-space volume expansion:
\begin{align}
W^{(N)}_{\mathrm{iso}}&\simeq -N_0\kb T\ln\left(\frac{\lambda_f}{\lambda_i}\right) ,\nonumber \\W_{\mathrm{cyc}}^{\mathrm{dil}} &\simeq N_0\kb (T_H-T_L)\ln\left(\frac{\lambda_>}{\lambda_<}\right).
\label{eq:W_dilute_cyc}
\end{align}
Conversely, in the quantum-degenerate regime ($\chi\gg 1$), $\Li_2(1-\e^{-\chi})\to\pi^2/6$. The response is constrained by the saturation of the bosonic occupation manifold, producing an algebraic work law:
\begin{align}
W^{(N)}_{\mathrm{iso}}&\simeq\frac{A\pi}{24}(\kb T)^2\left(\frac{1}{\lambda_f}-\frac{1}{\lambda_i}\right), \nonumber \\
W_{\mathrm{cyc}}^{\mathrm{deg}} &\simeq \frac{A\pi}{24}\kb^2 (T_H^2-T_L^2)\left(\frac{1}{\lambda_<}-\frac{1}{\lambda_>}\right).
\label{eq:W_degenerate_cyc}
\end{align}

With ideal regeneration during the fixed-coupling thermalization branches, the cycle efficiency achieves the reversible Stirling bound $\eta_{\mathrm{rev,reg}}=1-T_L/T_H$. Therefore, the TDOS-driven crossover does not alter the fundamental efficiency limit; rather, it dictates the scaling of the useful work extracted. Finally, while these expressions establish reversible thermodynamic limits, practical engine designs must also weigh the physical cost, precision, and dissipation associated with specific control channels, such as tuning magnetic anisotropy versus Dzyaloshinskii--Moriya interactions.

\section{Magnonic systems: Observables and honeycomb application}
\label{sec:magnons}

Magnonic systems provide a natural platform for the present framework because their excitation spectra can be continuously reshaped through magnetic fields, anisotropies, exchange couplings, strain, and spin-orbit-induced Dzyaloshinskii--Moriya interactions (DMI). The coupling-response TDOS can be accessed through two thermodynamically equivalent routes. The first relies on entropy reconstruction via coupling-dependent calorimetry,
\begin{equation}
S(T,\lambda)=S(T_0,\lambda)+\int_{T_0}^{T}\frac{C_\lambda(T',\lambda)}{T'}\dd T',
\label{eq:S_calorimetry}
\end{equation}
followed by differentiation with respect to $\lambda$ at a fixed temperature and chemical potential. The second route utilizes the generalized force. Since the magnon Hamiltonian typically contains a coupling-dependent contribution $\mathcal{H}_{\text{SW}}(\lambda)$, one finds
\begin{equation}
X_\lambda=-\left\langle\frac{\partial\mathcal{H}_{\text{SW}}}{\partial\lambda}\right\rangle,
\qquad
\G_\lambda=\left(\frac{\partial X_\lambda}{\partial T}\right)_{\mu,\lambda}.
\label{eq:force_measurement}
\end{equation}
This establishes the operational content of the Maxwell relation: the TDOS response is the temperature derivative of the force conjugate to the spectral control. 

This macroscopic force is directly linked to the spectral formulation. For a quadratic spin-wave Hamiltonian $\mathcal{H}_{\rm SW}=\sum_{n,\mathbf{k}}\epsilon_{n\mathbf{k}}(\lambda)\gamma^\dagger_{n\mathbf{k}}\gamma_{n\mathbf{k}}$, where $\gamma_{n\mathbf{k}}^{\dagger}$ creates a magnon in the $n$th branch, the generalized force evaluates to
\begin{equation}
X_\lambda=-\sum_{n,\mathbf{k}}\frac{\partial \epsilon_{n\mathbf{k}}}{\partial\lambda}\left\langle\gamma^\dagger_{n\mathbf{k}}\gamma_{n\mathbf{k}}\right\rangle=-\sum_{n,\mathbf{k}}\frac{\partial \epsilon_{n\mathbf{k}}}{\partial\lambda}n_B\!\left(\epsilon_{n\mathbf{k}}-\mu\right).
\label{eq:Xlambda_observables}
\end{equation}
Consequently, the caloric response is governed by the thermal magnon force,
\begin{equation}
X_\lambda^{\rm th} = - \sum_{n,\mathbf{k}} \frac{\partial \epsilon_{n\mathbf{k}}}{\partial\lambda} n_B\left(\epsilon_{n\mathbf{k}}-\mu\right).
\label{eq:Xlambda_th_observables}
\end{equation}
The generalized-force route therefore measures how the occupation of thermally accessible modes evolves under spectral deformation, weighting each mode by $\partial\epsilon_{n\mathbf{k}}/\partial\lambda$.

While equilibrium magnons strictly require $\mu=0$, a fixed-$N$ protocol becomes relevant in driven gases where a quasi-equilibrium chemical potential is established by microwave pumping or spin injection \cite{Demokritov2006,Bozhko2016,Divinskiy2021,Cornelissen2015,Cornelissen2016}. Here, separating the fixed-$\mu$ DOS channel from the implicit-$\mu$ channel distinguishes pure spectral reshaping from the population-conserving redistribution of magnons. It is important to note, however, that $\G_\lambda$ remains a reversible quasi-equilibrium susceptibility; assessing practical engine superiority requires further device-level analysis of dissipation, hysteresis, and control precision.

To illustrate how this response distinguishes inequivalent low-energy DOS structures, we apply the framework to a two-dimensional ferromagnetic honeycomb model:
\begin{equation}
\mathcal{H}=-J\sum_{\langle i,j\rangle}\mathbf S_i\cdot\mathbf S_j-K_u\sum_i(S_i^z)^2+D\sum_{\langle\langle i,j\rangle\rangle}\nu_{ij}\hat{\mathbf z}\cdot(\mathbf S_i\times\mathbf S_j),
\label{eq:honeycomb_H}
\end{equation}
where $J>0$ is the exchange, $K_u>0$ is an easy-axis anisotropy, and $D$ is the next-nearest-neighbor DMI \cite{Kim2016,Owerre2016}. 

Near the Brillouin zone center ($\Gamma$), the lower magnon branch is approximately quadratic, $\omega_\Gamma(k)\simeq\Delta_\Gamma+\rho_s k^2$, yielding the local DOS $g_\Gamma(\epsilon)=A/(4\pi\rho_s)\Theta(\epsilon-\Delta_\Gamma)$. Modulating the spin-wave stiffness $\rho_s$ at a fixed gap produces $\G^\Gamma_{\rho_s}=-S_\Gamma/\rho_s$. If a general control $\lambda$ alters both the stiffness and the zone-center threshold, the response is
\begin{equation}
\G^\Gamma_\lambda=-\frac{\partial_\lambda\rho_s}{\rho_s}S_\Gamma-\frac{A\kb}{4\pi\rho_s}(\partial_\lambda\Delta_\Gamma)s_B(\Delta_\Gamma;T,\mu).
\label{eq:Gamma_general_response}
\end{equation}
This confirms that a moving global threshold generates a localized edge term.

A fundamentally different structure governs the inequivalent valley points $K$ and $K'$. A massless local Dirac cone centered at energy $\omega_K$ has a DOS $g_D(\epsilon)=\nu_D A/(2\pi v_D^2)|\epsilon-\omega_K|$. Modulating the Dirac velocity yields a response $\G^D_{v_D}=-2S_D/v_D$, dictated by the scaling exponent $d/\alpha=2$. Introducing a DMI-induced Dirac gap ($m_D$) modifies the dispersion to $E_\pm(q)=\pm\sqrt{v_D^2q^2+m_D^2}$, giving the branch-resolved DOS
\begin{equation}
g_D(\epsilon)=\frac{\nu_D A}{2\pi v_D^2}|\epsilon-\omega_K|\Theta(|\epsilon-\omega_K|-|m_D|).
\label{eq:Dirac_DOS_massive}
\end{equation}
Crucially, changing $m_D$ creates finite-energy band-edge contributions, not a global threshold shift. A zone-center gap alters the baseline thermal population, whereas a DMI-induced Dirac gap reconfigures the Berry curvature distribution. Thus, DMI tuning provides measurable variations in the magnon thermal Hall conductivity $\kappa_{xy}$, distinguishing topological transport changes from global caloric threshold shifts.

To estimate the scale of the response, consider representative cryogenic parameters: $T=10~{\rm K}$ ($k_B T\simeq 0.862~{\rm meV}$), $\Delta_\Gamma=1~{\rm meV}$, and $\rho_s=100~{\rm meV}\cdot\text{\AA}^2$. For $\mu=0$, the entropy density is
\begin{equation}
\frac{S_\Gamma}{A} = \frac{k_B}{4\pi\rho_s} \left[ \Delta_\Gamma {\rm Li}_1(z_\Gamma) + 2k_BT \, {\rm Li}_2(z_\Gamma) \right], \quad z_\Gamma = e^{-\Delta_\Gamma/k_BT}.
\label{eq:Gamma_entropy_estimate}
\end{equation}
With $z_\Gamma\simeq 0.31$, ${\rm Li}_1(z_\Gamma)\simeq 0.38$, and ${\rm Li}_2(z_\Gamma)\simeq 0.34$, the entropy density is $S_\Gamma/A \simeq 7.7\times10^{-4} \, k_B \, \text{\AA}^{-2}$. The stiffness-controlled response density is therefore
\begin{equation}
\frac{\G^\Gamma_{\rho_s}}{A} = -\frac{1}{\rho_s}\frac{S_\Gamma}{A} \simeq -7.7\times10^{-6} \, k_B \, {\rm meV}^{-1}\text{\AA}^{-4},
\label{eq:Gamma_TDOS_number}
\end{equation}
confirming that a controlled change of the stiffness produces an entropy shift governed directly by the parameter-resolved TDOS.

Experimentally, this framework can be probed across several complementary techniques. The Maxwell relation gives direct access to $\G_B=(\partial M/\partial T)_B$ for magnetic-field control, while DMI forces relate to vector spin chirality. Brillouin light scattering (BLS) can probe the redistribution of magnon occupations under external driving \cite{Serga2010,Demokritov2006,Divinskiy2021}. Inelastic neutron scattering (INS) resolves magnon-band gaps and velocities \cite{Chen2018,Cai2021}, and spin-Seebeck or nonlocal magnon-transport measurements provide access to the effective chemical potentials governing the fixed-$N$ redistribution channel \cite{Uchida2010,Cornelissen2015,Cornelissen2016}.
{ 
Practical control selection requires connecting a Hamiltonian parameter
to an experimentally calibrated actuator. A field, stiffness, or
anisotropy protocol can involve different accessible ranges and may
change several spectral parameters simultaneously. If an actuator
\(u\) modifies parameters \(\{\lambda_i\}\), its isothermal response at
fixed chemical potential is
\[
\mathcal G_u
=
\sum_i
\left(\frac{\partial S}{\partial\lambda_i}\right)_{
T,\mu,\{\lambda_{j\neq i}\}}
\frac{d\lambda_i}{du}.
\]
The stiffness and gap contributions in Eq.~\eqref{eq:Gamma_general_response} provide an example.
Thus, independently varying a model parameter and operating a physical
actuator need not define the same thermodynamic protocol.

Comparisons should use attainable entropy changes and reversible heat
exchanges over specified parameter excursions, rather than the raw
magnitudes of derivatives expressed in different units. Control
precision is also relevant: greater entropy sensitivity can facilitate
detection while increasing the effect of parameter uncertainty.
The reversible description requires sufficient equilibration and
maintenance of the assumed particle constraint.

Finally, reversible work exchanged with the medium and energy consumed
by the control apparatus are distinct quantities. The coupling response
determines the thermal variation of the generalized force; an absolute
force reference or thermodynamic potential is needed for the work of an
individual stroke. Device-level comparisons additionally require
information about relaxation, hysteresis, dissipation, and actuation
energy. The present framework provides the equilibrium response for
such comparisons without establishing a universal ranking of controls.}

\section{Conclusions}
\label{sec:conclusion}

We have formulated an operational thermodynamic density of states (TDOS), $\G_\lambda=(\partial S/\partial\lambda)_{T,\mu}$, for bosonic working media controlled by spectral parameters rather than chemical potentials. For noninteracting bosons at fixed $\mu$, the response is strictly DOS-driven. Homogeneous gapped spectra, $\epsilon=\Delta+\lambda k^\alpha$, obey the exact scaling law $\G_\lambda=-(d/\alpha)S/\lambda$. Coupling-dependent gaps generate localized edge terms only when the spectral threshold possesses a finite boundary DOS weight, distinguishing true global thermodynamic edges from finite-energy interband crossings.

Under a fixed-particle constraint, $\mu(T,\lambda)$ becomes an implicit function, activating a secondary population-redistribution channel. In homogeneous spectra, this prevents a finite-density critical divergence, leading instead to a bounded normal-phase threshold response. In confined harmonic traps across the Bose--Einstein condensation transition, the dimensionality strictly dictates the analytical structure: the fixed-$N$ TDOS exhibits a finite jump at $T_c$ in three dimensions, but remains continuous in two dimensions. Furthermore, for pure scaling parameters, this constrained response is directly mapped to the heat capacity via the exact identity $\lambda\G_\lambda^{(N)}=-C_N$.

This TDOS fundamentally governs the reversible isothermal work in Stirling-type bosonic cycles, dictating a crossover from a logarithmic expansion law in the classical dilute limit to an algebraic dependence in the quantum-degenerate regime. We connected this formalism to observable generalized forces in magnonic platforms, demonstrating how coupling-dependent calorimetry and force measurements can resolve structural distinctions—such as separating the global quadratic zone-center threshold from DMI-induced topological Dirac gaps in honeycomb ferromagnets.

Ultimately, this framework establishes a unified thermodynamic language for coupling-controlled bosonic media. By identifying how spectral reshaping, dimensionality, threshold structure, and particle-number constraints govern entropy and work, the coupling-response TDOS provides direct operational tools for evaluating energy costs and heat exchanges in systems ranging from solid-state magnons and cold atoms to microwave cavity modes in superconducting circuits and quantum batteries.
{ 
Experimentally, these operational predictions can be tested by tracking
entropy and reversible heat along calibrated control paths. Quasistatic
heat-engine strokes offer one such test: the predicted reversible heat
exchange can be compared with measured cycle data \cite{Cruz_2023}. In
magnonic media, the temperature dependence of magnetization provides a direct
Maxwell-relation test, complemented by coupling-dependent calorimetry
once background contributions are separated \cite{10.1063/5.0011887}. In cold-atom traps,
frequency variation at nearly fixed particle number probes the
constrained response \cite{WOS:001493202700001}; under pure spectral scaling,
\(\lambda\mathcal G_\lambda^{(N)}=-C_{\lambda,N}\), which can be
compared with heat-capacity measurements. Microwave cavity modes in
superconducting circuits offer another setting once their spectra,
control paths, and state preparation are characterized. Related
cavity-QED and optomechanical platforms can be explored under the same
equilibrium or quasi-equilibrium requirement
\cite{WOS:000472681100013,WOS:001840550200009}. Quantum batteries offer
a further direction, provided their state constraints and the
applicability of an equilibrium or quasi-equilibrium description are
specified.
Such measurements would test the framework's equilibrium predictions,
while finite-time dynamics, imperfect thermalization, and actuator
losses remain questions for further study.}

\section*{Acknowledgments}
F.J.P., B.C., and N.V.-S. acknowledge support from FONDECYT (Chile) under Grant No.~1250173. N.V.-S. acknowledge funding from FONDECYT Regular Grant  No.~1250364, respectively. B.C. acknowledge support from PUCV and  Direcci\'on de Postgrado of UTFSM. B.C. further acknowledges support from the Programa de Incentivo a la Iniciaci\'on Cient\'ifica (PIIC) No.~005/2026. B.C. acknowledges financial support from the Agencia Nacional de Investigación y Desarrollo (ANID), Chile, through the National Doctoral Scholarship and the Complementary Benefits for Operational Expenses [Grant/Folio No. 21250015].

\end{document}